\documentclass[iop, twocolumn, tighten]{emulateapj}
\usepackage{amsmath}
\usepackage{color}
\newcommand{\omegaspin}{\omega_{\mathrm{spin}}}

\newcommand{\vrrot}{v_{r,\mathrm{rot}}}
\newcommand{\vrorb}{v_{r,\mathrm{orb}}}
\newcommand{\vrp}{V_r}
\newcommand{\Krot}{K_\mathrm{rot}}
\newcommand{\vbarrot}{\bar{v}_{r,\mathrm{rot}}}
\newcommand{\lamp}{\lambda_p}

\newcommand{\ipl}{i_p}
\newcommand{\hatTheta}{\hat{\Theta}}
\newcommand{\er}{{\boldsymbol e}_R}
\newcommand{\elos}{{\boldsymbol e}_\mathrm{O}}
\newcommand{\es}{{\boldsymbol e}_\mathrm{S}}
\newcommand{\IRR}{\mathrm{IRR}}
\newcommand{\DR}{\mathrm{DR}}

\slugcomment{Accepted for publication in The Astrophysical Journal Letters}
\shortauthors{Kawahara}
\shorttitle{Spin Effect on Planetary Radial Velocity}
\begin{document}
\title{The Spin Effect on Planetary Radial Velocimetry of Exoplanets}

\author{Hajime Kawahara\altaffilmark{1}} 
\altaffiltext{1}{Department of Physics, Tokyo Metropolitan University,
  Hachioji, Tokyo 192-0397, Japan}
\email{kawa\_h@tmu.ac.jp}
\begin{abstract}

We consider the effect of planetary spin on the planetary radial velocity (PRV) in dayside spectra of exoplanets. To understand the spin effect qualitatively, we derive an analytic formula of the intensity-weighted radial velocity from planetary surface on the following assumptions: 1) constant and solid rotation without precession, 2) stable and uniform distribution of molecules/atoms , 3) emission models from dayside hemisphere, and 4) a circular orbit. On these assumptions, we find that the curve of the PRV is distorted by the planetary spin and this anomaly is characterized by spin radial velocity at equator and a projected angle on a celestial plane between the spin axis and the axis of orbital motion $\lambda_p$ in a manner analogous to the Rossiter-McLaughlin effect. The latter can constrain the planetary obliquity. Creating mock PRV data with 3 km/s accuracy, we demonstrate how $\lambda_p$ and the spin radial velocity at equator are estimated. We find that the stringent constraint of eccentricity is crucial to detect the spin effect. Though our formula is still qualitative, we conclude that the PRV in the dayside spectra will be a powerful means for constraining the planetary spin. 
\end{abstract}
\keywords{planets and satellites: fundamental parameters -- techniques: radial velocities -- techniques: spectroscopic}


\section{Introduction}

Planetary spin is one of crucial factors that govern the climate of exoplanets \citep[e.g.][]{1997Icar..129..254W, 2003IJAsB...2....1W, 2011A&A...528A..27H,2012arXiv1205.5034C} and has a potential to constrain planetary formation theory \citep[e.g.][]{1999Icar..142..219A,2001Icar..152..205C,2007ApJ...671.2082K}. Photometric variation of a planet will enable us to estimate the rotation period \citep{2008ApJ...676.1319P} and the obliquity \citep{2010ApJ...720.1333K,2011ApJ...739L..62K,2012ApJ...755..101F} in the near future. However these methods are applicable only for a planet having significant inhomogeneous surface, such as coexistence of ocean and lands or clouds, and also require a long-term observation with a sophisticated instrument of direct imaging.  

Recently {\it planetary} radial velocity (PRV) of the non-transiting planet, $\tau$ Bo{\"o}tis b has been measured with the aid of the carbon monoxide absorption in the thermal dayside spectrum \citep{brogi,2012ApJ...753L..25R}. \cite{brogi} detected the change in the radial component of the planet's orbital motion and obtained the semiamplitude of the PRV, $K_p=110.0 \, \pm \, 3.2$ km/s. Combining the {\it stellar} radial velocimetry with it, they evaluated the orbital inclination and mass of $\tau$ Bo{\"o}tis b. Though the PRV curve is primarily dictated by the planet's orbital motion, the planetary spin also has possible effect on the PRV in the dayside spectrum.  Gas giants in the solar system have considerable spin velocity at equator, 12 km/s for Jupiter and 10 km/s for Saturn. Even for most hot Jupiters which are likely to be in a synchronous orbit, are expected to have non negligible spin velocity driven by their rapid orbital motion. For instance, WASP 19b will have $\sim 9$ km/s of the spin velocity at equator if it is tidally locked.  
The aim of the paper is to develop the method to derive the {\it planetary} spin-orbit alignment and the spin velocity from time series analysis of the PRV. This concept can be explained by an analogy to the Rossiter-McLaughlin effect \citep[RM effect][and references therein]{2000A&A...359L..13Q,2005ApJ...622.1118O,2005ApJ...631.1215W,2007ApJ...655..550G}. The RM effect is an anomaly of {\it Stellar} radial velocity (SRV) caused by sequent occultation of a rotating stellar disk by a transiting planet, and is used to measure the projected angle of the orbital axis and the {\it stellar} spin axis \citep{2005ApJ...622.1118O}. Likewise, non-uniform emission from a planet, which is generally stronger near the sub-stellar direction, induces an anomaly in a time series of the PRV. In this paper, we demonstrate how the planetary spin affects the PRV assuming a simple solid rotation of a planet and derive an analytic formula of the PRV anomaly with simple intensity distribution models. 

\section{Methods}
\begin{figure*}
\begin{center}
  \includegraphics[width=\linewidth]{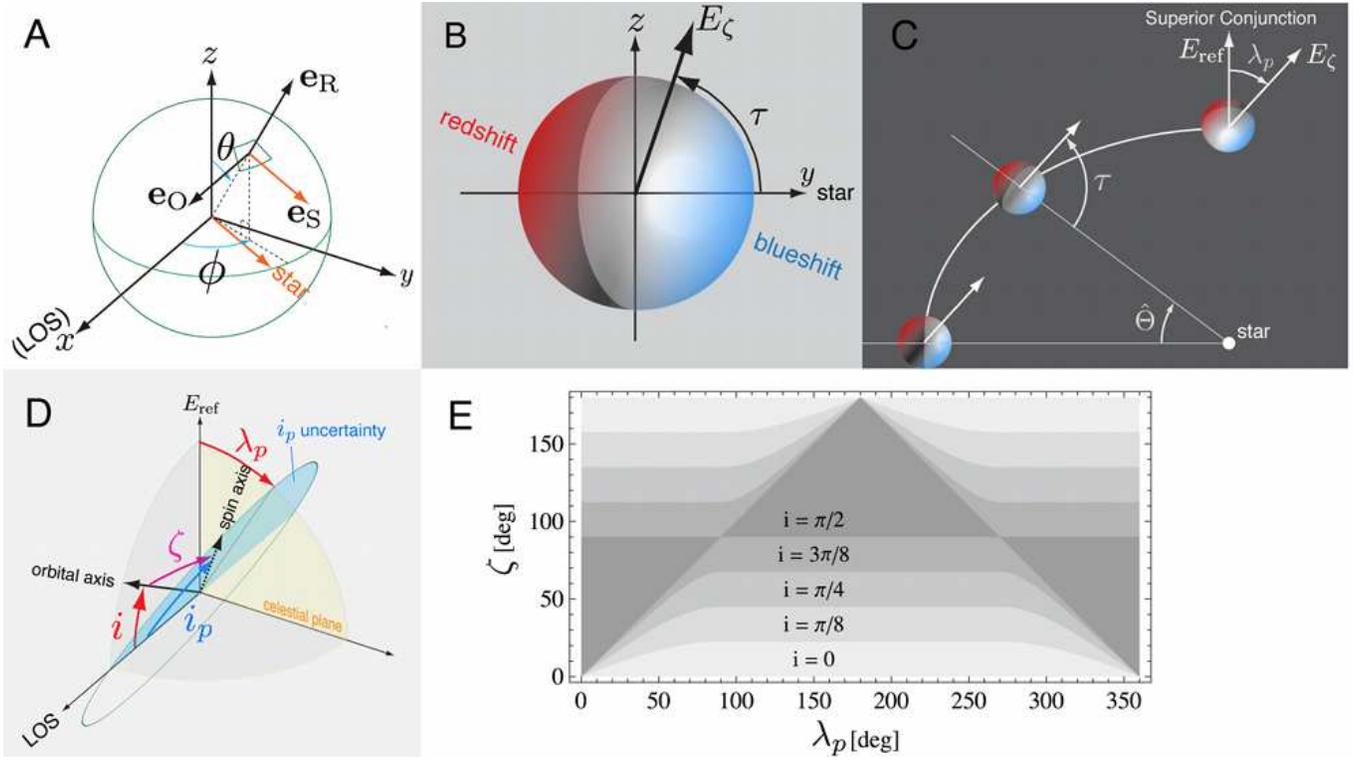}
\caption{Geometric configuration. Panel A: the spherical coordinates on a planet. The stellar direction is embedded in the $x$-$y$ plane. The unit vectors $\elos$, $\es$, and, $\er$ are explained in the text. Panel B and C presents schematic pictures from observer's point of view ($y$-$z$ plane). In panel B, the blue shifted emission dominates on the dayside hemisphere (drawn by white), which causes the blue-shifted anomaly in the PRV. The planetary spin orbit misalignment $\lambda_p$ is defined as the projected angle between the spin axis and the orbital axis on the celestial plane (Panel C).  Panel D indicates relation of the angles, the obliquity $\zeta$, the orbital inclination $i$, the spin inclination $i_p$, and $\lambda_p$. The uncertainty range of $\ipl$ due to the degeneracy between $\sin{\ipl}$ and $\omegaspin R_p$ for a solid rotator is shown. Panel E: possible range of obliquity $\zeta$ as a function of $\lamp$ (Eq. [\ref{eq:prog}]). Each shade corresponds to different orbital inclination $i$.
 \label{fig:geo}}
\end{center}
\end{figure*}

We divide the PRV into the radial velocity components of the planetary center system $\vrorb(\Theta)$ and the planetary spin $\vrrot (\Theta)$,
\begin{eqnarray}
v_r(\Theta) &=& \vrorb (\Theta) + \vrrot (\Theta). 
\end{eqnarray}
On the assumption of a circular orbit for simplicity, the PRV by the orbital motion is expressed as   
\begin{eqnarray}
\vrorb (\Theta) &=& K_p \cos{\Theta},
\end{eqnarray}
where $K_p$ is semiamplitude of the radial velocity by the orbital motion and $\Theta$ is the orbital phase.  We define the phase angle between the line of sight and star-planet direction $\alpha$,
\begin{eqnarray}
\cos{\alpha} &=& \es \cdot \elos = \sin{i} \sin{\Theta} \nonumber \\  
&\,& \mbox{( $\pi/2 - i \le \alpha \le \pi/2 + i$ )},
\end{eqnarray}
where $\es = (\cos{\alpha}, \sin{\alpha}, 0)^T$ and $\elos = (1,0,0)^T$ are unit vectors from the planetary center to the star and the observer and $i$ is the orbital inclination (Figure \ref{fig:geo} A).

Since the observed absorption consists of ensemble of a Doppler-shifted line from each position on planetary surface, $\vrrot (\Theta)$ depends on the distribution of absorption lines on the surface. We use the intensity distribution instead of the distribution of absorption lines assuming the uniform distribution of molecules. \cite{2005ApJ...622.1118O} showed that the intensity-weighted velocity is in agreement with the Doppler shift, $\vbarrot/c = \Delta \nu/\nu$ on the assumption that the frequency shift is much smaller than the line frequency (Eq. [20] in their paper). Though the precise value measured by real observation will depends on details of methods and instruments, the aim of the paper is to qualitatively understand the behavior of $\vrrot (\Theta)$. Hence we regard the intensity-weighted velocity,  
\begin{eqnarray}
\label{eq:weighted}
\vbarrot &\equiv& \frac{1}{\Psi(\alpha)}\int d \Omega \, W(\phi,\theta; \alpha) \, \vrp(\phi,\theta) \equiv \langle  \vrp \rangle, \\
\Psi(\alpha) &\equiv& \int d \Omega \, W(\phi,\theta; \alpha) 
\end{eqnarray}
as the PRV of the spin, $\vrrot (\Theta)$, where the $\Psi (\alpha)$ is the phase function (total intensity of emission), $W(\phi,\theta; \alpha)$ is the intensity distribution normalized so that $\Psi(0)$ is unity, $\vrp (\phi,\theta)$ is the radial velocity of a facet on surface measured on the planetary center system, and  $d \Omega = \sin{\theta} d \theta \, d \phi $ (see Fig. \ref{fig:geo}A).  

A naive expectation is that $W(\phi,\theta; \alpha)$ has higher value at the dayside hemisphere than that at the nightside hemisphere (Fig \ref{fig:geo} B and C). This is absolutely the case for the scattered light. It is also likely to be the case for thermal emission on the assumptions that the energy injection is mainly attributed to incident light from the host star.  

In this paper, we regard a planet as a solid rotator and no precession,
\begin{eqnarray}
\label{eq:vrporch}
\vrp (\phi,\theta; \tau) = \Krot (-\sin{\tau} \sin{\phi} \sin{\theta} + \cos {\tau} \cos {\theta} ),
\end{eqnarray}
where $\tau \equiv \hatTheta + \pi/2 - \lamp$ is the projected rotation angle between the spin axis and the stellar direction on celestial plane and the projected orbital phase angle $\hatTheta$ is expressed as
\begin{eqnarray}
\tan{\hatTheta} = \cos{i} \, \tan{\Theta}.
\label{eq:hatt}
\end{eqnarray}
The maximum radial velocity of the solid rotator $\Krot$ represents the strength of the spin,
\begin{eqnarray}
\Krot \equiv 2 \pi \omegaspin R_p \sin{\ipl},
\end{eqnarray}
where $\ipl$ is the spin inclination. Thus the degeneracy between $\sin{\ipl}$ and $\omegaspin$ is inevitable for a solid rotator. The planetary obliquity $\zeta$ is expressed as,
\begin{eqnarray}
\cos{\zeta} = \cos{i} \cos{\ipl} + \sin{i} \sin{\ipl} \cos{\lamp},
\label{eq:obl}
\end{eqnarray}
where we introduce the planetary spin orbit misalignment $\lambda_p$, defined by the projected angle between the spin axis and the orbital axis on the celestial plane. Figure \ref{fig:geo} D displays a schematic picture of these angles. Though $\ipl$ is difficult to know directly, if knowing $\lamp$ from the PRV time series analysis, one can constrain the obliquity $\zeta$ for a given $i$. Figure \ref{fig:geo} E shows the possible region of $\zeta$ for given $\lamp$,
\begin{eqnarray}
\sin^{-1}{(\sin{i} |\sin{\lambda_p}|)} \le \zeta \le \pi - i \nonumber
\end{eqnarray}
for $0 \le \lamp < \pi/2$, $3\pi/2 \le \lamp < 2 \pi$, 
\begin{eqnarray}
i \le \zeta \le \pi - \sin^{-1} {(\sin{i} |\sin{\lambda_p}|)} 
\label{eq:prog}
\end{eqnarray}
for  $\pi/2 \le \lamp < 3 \pi/2$.
\begin{figure}
  \includegraphics[width=\linewidth]{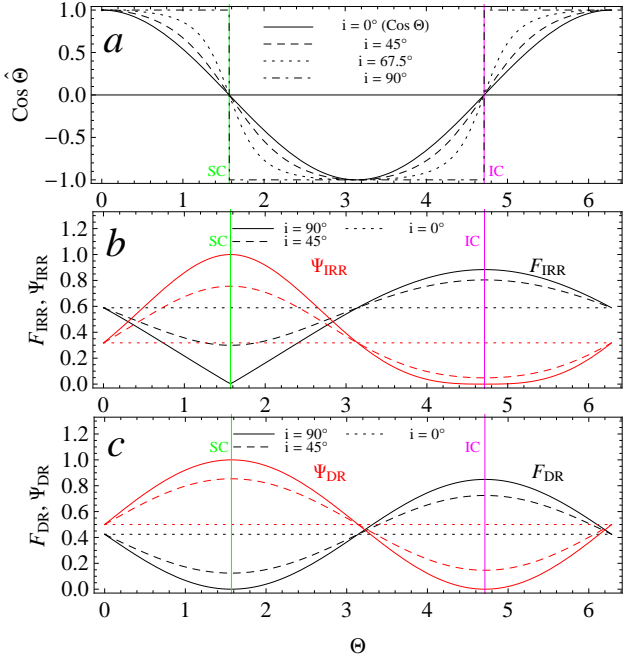}
\caption{  Panel a: $\cos{\hatTheta}$ as a function of $\Theta$. Panels b and c: $F(\Theta)$ (black) and phase function $\Psi(\Theta)$ (red) for the IRR and DR models. Two vertical lines correspond to SC and IC. \label{fig:geobehave}}
\end{figure}

If the intensity distribution is symmetric about the stellar direction ($W(\phi, \theta; \alpha) = W(\phi, \pi/2 - \theta; \alpha)$), the antisymmetric $\cos{\theta}$ term in equation (\ref{eq:vrporch}) vanishes and equation (\ref{eq:weighted}) reduces to
\begin{eqnarray}
\vbarrot = - \Krot F(\alpha) \cos{(\hatTheta+\lamp)},
\label{eq:weightedsym}
\end{eqnarray}
where
\begin{eqnarray}
\label{eq:falpha}
F(\alpha) \equiv \frac{1}{\Psi(\alpha)} \int d \Omega \sin{\theta} \sin{\phi} \, W(\phi,\theta; \alpha). 
\end{eqnarray}
The $\phi$-symmetric component of $W(\phi,\theta; \alpha)$ does not contribute to $\vrrot$. In equation (\ref{eq:weightedsym}), one can interpret that $F(\alpha)$ represents the inhomogeneous pattern of the planetary surface. The $\cos{(\hatTheta+\lamp)}$ term is due to apparent change of dayside along orbit and does not depend on the intensity distribution. As increasing $i$, the deviation by this term from the cosine curve of the orbital motion becomes stronger as shown in Figure \ref{fig:geobehave} a.

\section{Analytic Formula of the spin effect with Radiation Models}

Considering two specific models of the intensity distribution, the instant re-radiation (IRR) model and the dayside re-distribution (DR) model \citep[e.g.][]{2007ApJ...667L.191L, 2011ApJ...726...82C,2012arXiv1203.6017S}, we derive an analytic expression of the spin effect. 

The IRR model assumes the input energy from a host star is instantly and isotopically emitted from the planetary surface. The isotropic reflection for the scattered light or no redistribution of energy around surface for the thermal emission satisfies this assumption. The intensity distribution of the IRR model is expressed as,
\begin{eqnarray}
\label{eq:lam}
W_\IRR(\phi,\theta) = \left\{
\begin{array}{c}
\displaystyle{\frac{3}{2 \pi} (\es \cdot \er) (\er \cdot \elos) } \\
 \mbox{\,\,\,\,\, for $-\pi/2+\alpha \le \phi \le \pi/2$,} \\
0  \mbox{\,\,\,\,\, for elsewhere, } \\
\end{array} \right.
\end{eqnarray}
and the phase function
\begin{eqnarray}
\Psi_\IRR(\alpha) = \frac{1}{\pi} (\sin{\alpha}+(\pi-\alpha ) \cos{\alpha}),
\end{eqnarray}
where $\er = (\cos{\phi} \sin{\theta},\sin{\phi} \sin{\theta},\cos{\theta})$. Using equation (\ref{eq:falpha}), we obtain
\begin{eqnarray}
F_\IRR(\alpha) &=&  \frac{3 \pi  \sin {\alpha} (\cos {\alpha}+1)}{16 [(\pi -\alpha ) \cos {\alpha}+\sin{\alpha}]}.
\end{eqnarray}
The DR model assumes rapid energy redistribution on the dayside.
\begin{eqnarray}
\label{eq:dr}
W_\DR(\phi,\theta) &=&
\left\{
\begin{array}{c}
\displaystyle{\frac{1}{\pi} (\er \cdot \elos) } \mbox{ for $-\pi/2+\alpha \le \phi \le \pi/2$,}  \\
0  \mbox{\,\,\,\,\, for elsewhere, }  \\
\end{array} \right. \\
\Psi_\DR(\alpha) &=& \cos^2 \left( {\frac{\alpha}{2}} \right), \\
F_\DR(\alpha) &=&  \frac{8}{3 \pi} \sin^2 \left( {\frac{\alpha}{2}} \right).
\end{eqnarray}

\begin{figure*}
  \includegraphics[width=0.49\linewidth]{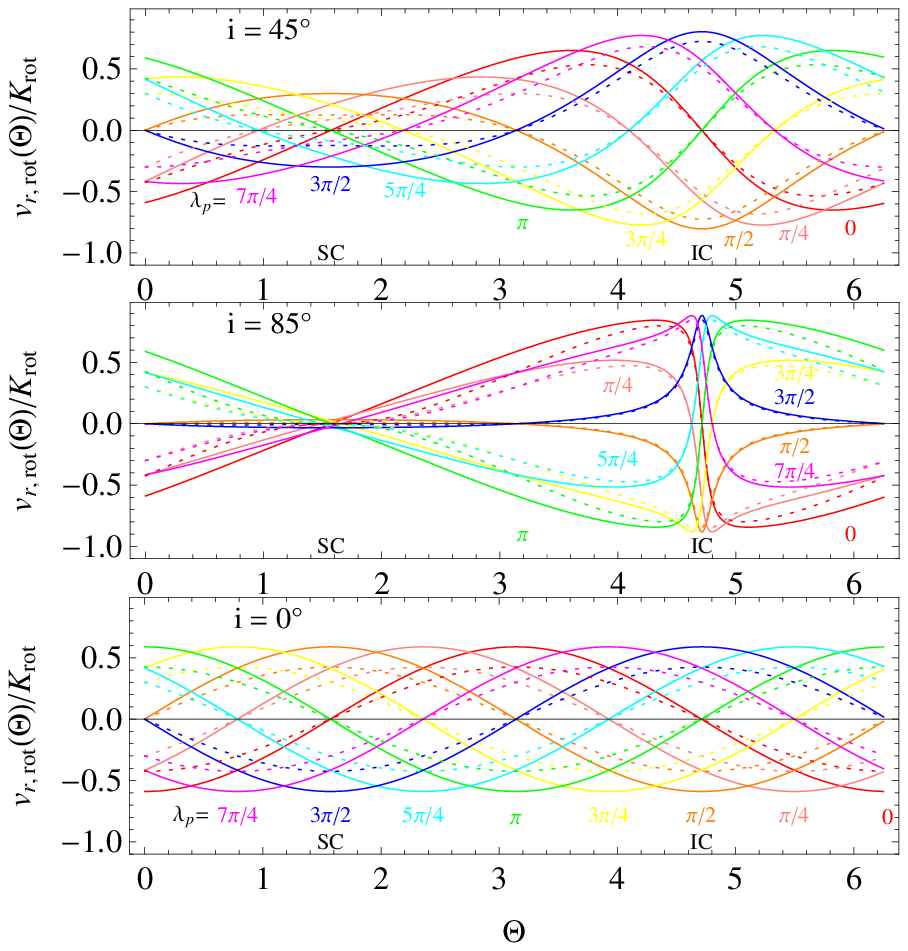}
  \includegraphics[width=0.49\linewidth]{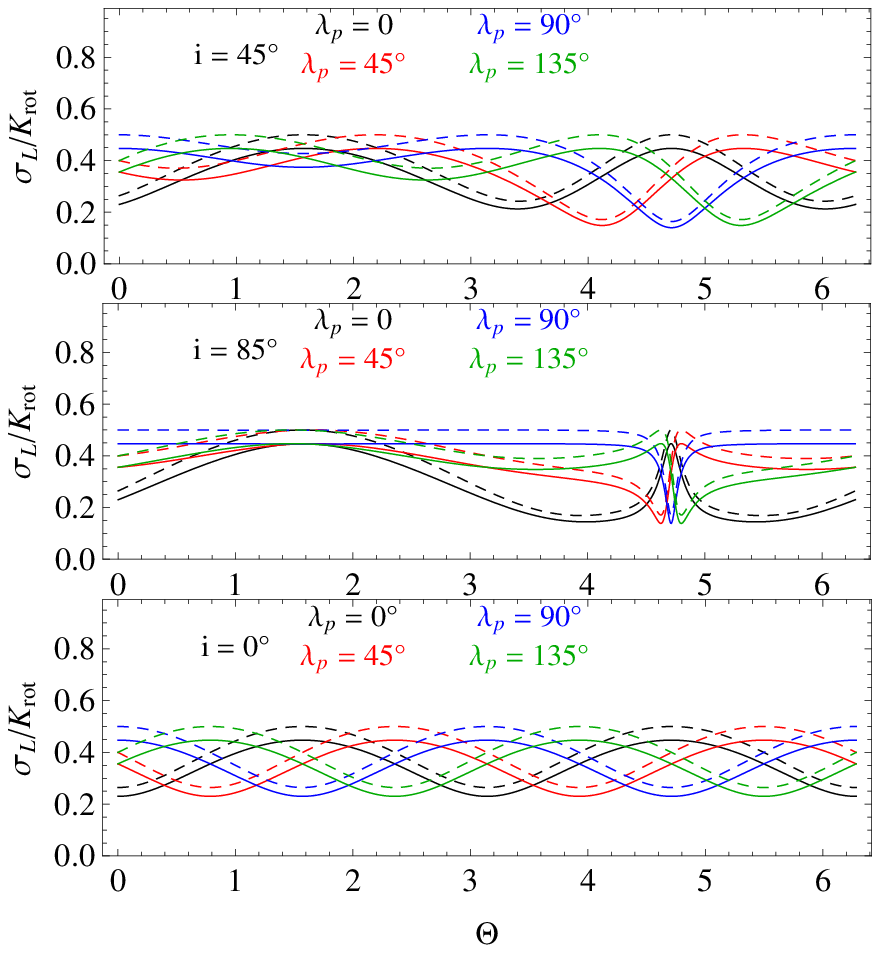}
\caption{Left panels show the PRV anomaly by planetary spin as a function of $\Theta$ in an inclined orbit ($i=45^\circ$; top) and a typical transiting system ($i=85^\circ$; middle), and a face-on orbit ($i=0^\circ$; bottom). Each color corresponds to different $\lamp$. Solid lines assumes the IRR model, while dotted lines indicate the DR model. Right: line broadening due to a solid rotation for different inclination (same as the right panel). Standard deviation normalized by $\Krot$ is shown by solid (dashed) lines for the IRR  (DR) model. \label{fig:vrrot}}
\end{figure*}

Figure \ref{fig:geobehave}b and c display $F$ and $\Psi$ for both models as a function of $\Theta$. Substituting $F_\IRR(\alpha)$ ($F_\DR(\alpha)$) into equation (\ref{eq:weightedsym}), we obtain the analytic expression of $\vbarrot$. Figure \ref{fig:vrrot} (left) shows  $\vbarrot$ for different $\lamp$ and $i$. One can see characteristic features of the PRV depending on $\lamp$ and $i$. By fitting this anomaly, we can estimate $\lamp$ and $\Krot$ as will be demonstrated in next section. The difference of $\vbarrot$ between these models is not significant.  

For the exact edge-on orbit, equation (\ref{eq:weightedsym}) reduces to 
\begin{eqnarray}
\vbarrot = \pm 2 \pi \omegaspin R_p \cos{\zeta} F(\alpha),
\label{eq:transit_v}
\end{eqnarray}
where a positive sign is for $\pi/2 \le \Theta \le 3 \pi/2$ and negative for $\Theta \le \pi/2$ or $\Theta \ge 3 \pi/2$. Equation (\ref{eq:transit_v}) indicates that the phase information of the term $\cos{(\hatTheta + \lamp)}$ disappears except for its sign  in the edge on limit. However one can know whether the spin is prograde or retrograde against the orbit.


While we have considered the line shift so far, the spin rotation also induces the line broadening. We simply estimate the doppler broadening by considering the variance,
\begin{eqnarray}
\label{eq:var}
\sigma_\mathrm{L}^2 \equiv \langle  \vrp^2 \rangle - \langle  \vrp \rangle^2,
\end{eqnarray}
where the definition of $\langle \rangle$ is same as that in equation (\ref{eq:weighted}). The right panels in Figure \ref{fig:vrrot} show $\sigma_\mathrm{L}$ for $i=45^\circ, 85^\circ$ and $0^\circ$. Typical broadening width is $\sim 0.4 \, \Krot$. The $\Theta$-dependence is smaller than that of the line shift. The dependence of the emission models on $\sigma_\mathrm{L}$ is not significant. If using extremely high dispersion spectroscopy (for instance, $R=300,000$ for $\Krot=10$ km/s), one may derive this characteristic feature of the broadening variation. 

\section{Demonstration}
\begin{figure*}
  \includegraphics[width=0.49 \linewidth]{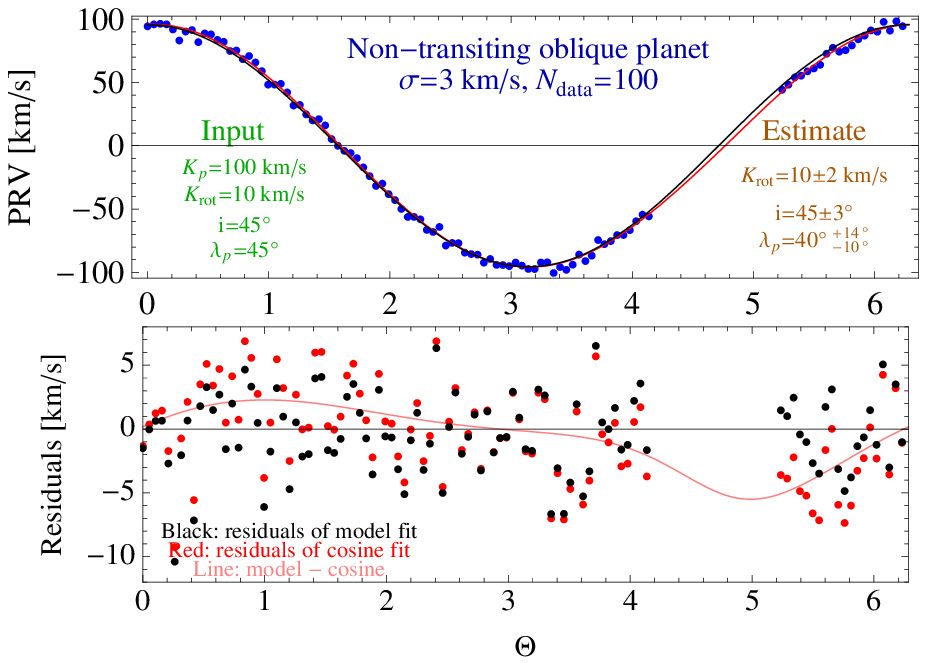}
  \includegraphics[width=0.49 \linewidth]{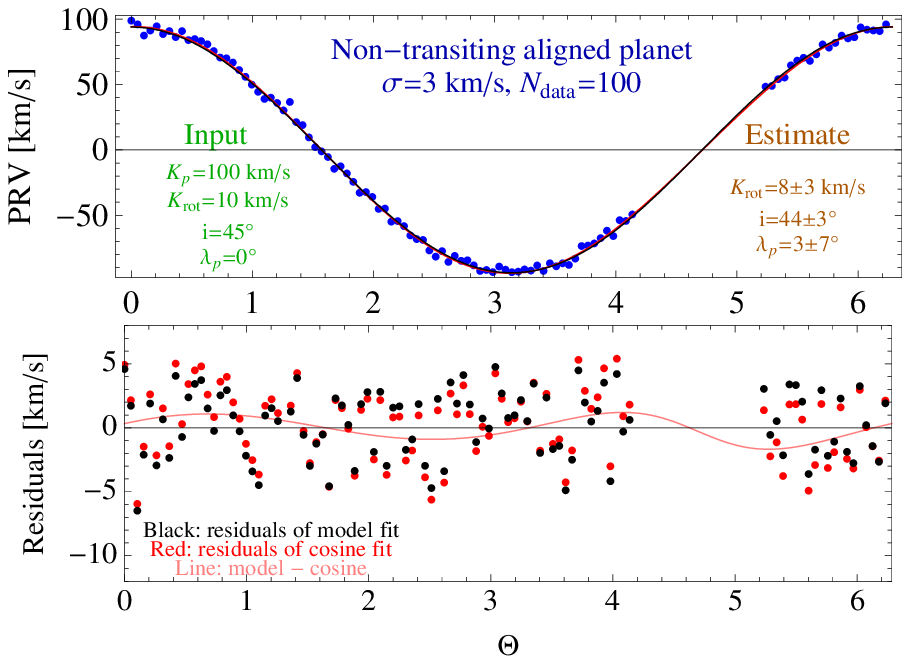}
  \includegraphics[width=0.49 \linewidth]{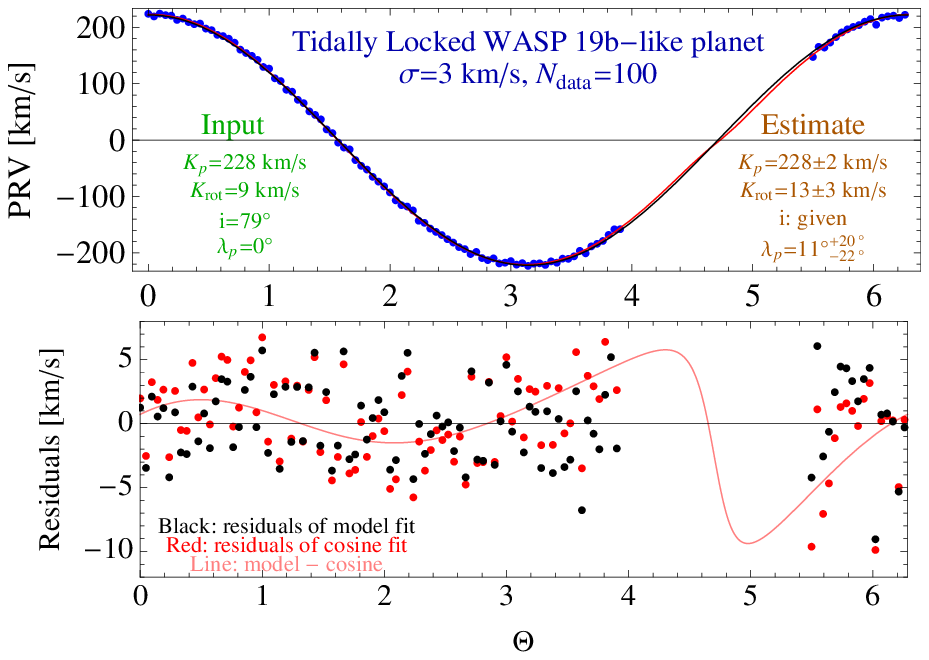}
  \includegraphics[width=0.49 \linewidth]{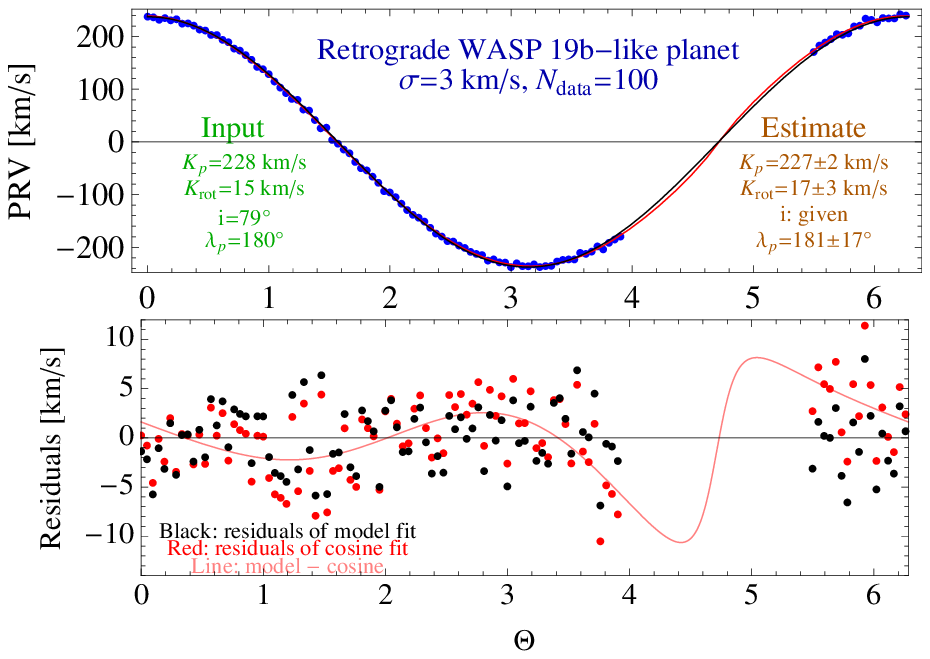}
\caption{Mock data of the PRV and residuals with 3 km/s accuracy. Top panels: mock PRV curves for oblique (left; $\lamp=45^\circ$) and aligned (right; $\lamp=0^\circ$) planets for a non-transiting system ($i=45^\circ$). Black and red curves in the upper subpanels are the model fit and cosine fit of the data. The bottom subpanels display the residual of the model fitting (black) and the cosine fitting (red). Solid curves indicate the model fitting minus the cosine fitting, that is, the anomaly from the cosine curve. Bottom panels: simulated data for a transiting system mocking the WASP 19 b-like system ($i=79^\circ$). Bottom left is a tidally locked case (the planet's rotation is directly driven by the orbital motion). Bottom right: an aligned planet but having the rapid retrograde spin ($\lamp=180^\circ$) with $\Krot=15$ km/s. \label{fig:examock}}
\end{figure*}

We demonstrate the spin effect by mocking the PRV time series. Figure \ref{fig:examock} shows the mock PRV curves with the IRR model and its fitting curves. We set $\sigma = 3$ km/s precision, roughly corresponding to the accuracy obtained when lines are barely detected and resolved with $R \sim 100,000$, though it can be improved by increasing sensitivity and the number of lines. The top panels in Figure \ref{fig:examock} assumes a non-transiting system with $i=45^\circ$, $K_p=100$ km/s, $\Krot=10$ km/s, and $\lamp=45^\circ$ (left; case A) or $0^\circ$ (right; case B). We take 100 data points and avoid the $\pm 30^\circ$ range around the inferior conjunction (IC) since the planetary signal in this range declines below 7 \% (20 \%)  of the maximum value at $\Theta=90^\circ$ for the IRR (DR) model (see Figure 2bc red lines). We use a Gaussian prior for the stellar mass with typical 5 \% uncertainty and simultaneously fit the stellar mass, $i$, $\lamp$, and $\Krot$ using a Markov chain Monte Carlo (MCMC) algorithm. The estimated inclination is well constrained $i=45 \pm 3^\circ$ since it is almost determined by the whole amplitude. The spin effect can be detected above 3 $\sigma$ for these cases; $\Krot=10 \pm 2$ km/s with $\lamp=40^{+14}_{-10} \,^\circ$ (case A) and $\Krot=8 \pm 3$ km/s with $\lamp=3 \pm 7 \,^\circ$  (case B).

Though we have assumed an exact circular orbit so far, uncertainty of small eccentricity, $e$ can be a possible source of systematics. For $e \ll 1$, one can express the velocity modulation due to $e$ as,
\begin{eqnarray}
\vrorb \approx K_p [\cos{\Theta} + e \cos{(2 \Theta - \omega)}],
\label{eq:approx}
\end{eqnarray}
where $\omega$ is the argument of periastron. The modulation has the amplitude of $e K_p$ and double frequency. We perform the MCMC leaving $e$ and $\omega$ as fitting parameters with equations (\ref{eq:weightedsym}) and (\ref{eq:approx}). Assuming a typical constraint of $e < 0.02$, we obtain slightly worse constraints, $\Krot=8^{+5}_{-3}$ km/s and $\lamp=45^{+40}_{-20} \,^\circ$ for the case A. For the case B, the eccentricity uncertainty makes the spin effect to be unconstrained (1 $\sigma$ detection level of $\Krot$) due to the correlation between $e$ and $\Krot$. Thus $e<0.02$ is marginal to detect the spin effect since the uncertainty of the modulation is comparable to the amplitude of the spin velocity, that is, $K_p \Delta e \sim \Krot$. If assuming $e < 0.002$, which satisfies $\Delta e \ll \Krot/K_p$, we obtain $\Krot = 9^{+3}_{-3}$ km/s and $2^{+6}_{-8} \,^\circ$ even for the case B. As shown in the top left panel, $\vrrot$ for the case B is resemble to the eccentricity modulation (Eq. [\ref{eq:approx}]) in its curve. Therefore the case B is more sensitive to the $e$ uncertainty than the case A. A more stringent constraint of $e$ than $\Krot/K_p$ is important to detect the spin effect. 
 
For the time being, a realistic target of the PRV measurement will be confined to hot Jupiters. We also consider the application of the precise PRV measurement to hot Jupiters. Orbital period of hot Jupiters is generally short ($P \sim$ a few days, typically). In a synchronous orbit, the rotation period equals to the orbital period. Contribution of the orbital motion to $\vrrot$ is $2 \pi R_p/P =1-3$  km/s for typical hot jupiters and reaches $2 \pi R_p/P =9 $  km/s for the extreme case, WASP 19b \citep[$P=1.2$ day and $R_p=1.39 R_J$;][]{2011ApJ...730L..31H}. We create the mock PRV of a tidally locked transiting planet with parameters of WASP 19b ($\Krot=228$ km/s and $i=79^\circ$). Since $i$ is known for the transiting planet, we assume 5 \% uncertainty of the stellar mass again. We fit the PRV curve avoiding the range $\pm 45^\circ$ around transit, in which the planetary signal declines below 5 \% (IRR) and 15 \% (DR) of the maximum. We also take the eccentricity uncertainty into consideration by adopting $e=0.046$ and $\omega=0^\circ$ for input. Since current uncertainty of WASP 19 b is $e = 0.0046^{0.0044}_{-0.0028}$ and $\omega = 3 \pm 70^\circ$ \citep{2011ApJ...730L..31H}, we assume a Gaussian prior with $\sigma_e = 0.004$ and $\sigma_\omega = 70^\circ$. We obtain a marginal detection of the spin, $\Krot = 11^{+6}_{-5} $ km/s and $\lamp = 4^{+28}_{-30} \,^\circ$ . The fitting with more stringent constraints $\sigma_e=0.001$ and $\sigma_\omega=20^\circ$ provides $\Krot = 13 \pm 3 $ km/s and $\lamp = 11^{+20}_{-22} \,^\circ$. The bottom left panel shows the mock PRV and its fitting results for the latter.

Most hot Jupiters are likely to be in a synchronous orbit with $\zeta=0$ (therefore $\lamp=0$) due to tidal force. However, this expectation has never been proved observationally. In particular, the assumption of the tidal lock is not obvious for a planet in a highly eccentric orbit such as HD 80606 b \citep[$e=0.93$;][]{2009MNRAS.396L..16F}. Finally we consider an extreme planet with a rapid retrograde spin $\zeta=180^\circ$ and $\Krot=15$ km/s. We note that { \it retrograde } here means the retrograde rotation of the { \it planetary } spin ( not the stellar spin ) against the orbital revolution, like Venus. Performing the same fitting process as the tidally locked case (with $\Delta e=0.004$ and $\Delta \omega = 70^\circ$), we obtain $\Krot=17 \pm 3$ km/s and $\lamp=181 \pm 17^\circ$ with a characteristic feature of the residual as shown in the bottom right panel.

In this section, we excluded the light curve around the IC where the signal is weaker as is clear from $\Psi(\Theta)$ in Figure \ref{fig:geobehave}. As shown in the bottom subpanels of Figure \ref{fig:examock}, most characteristic features appear around the IC though it depends on $i$ (see also Figure \ref{fig:vrrot}). This is one of main difficulties to detect the spin effect practically. Though current detection of the PRV is far away from the IC \citep[$0.5 < \Theta < 2.5$ for ][]{brogi}, future detection in the outer range is of importance to detect the spin effect.

\section{Discussion and Conclusions}

In this paper, we have estimated the observable shift of the planetary absorption by substituting the intensity weighted radial velocity (Eq. [\ref{eq:weighted}]). It is known that the cross-correlation method used in the RM effect has a few - 50\% systematics of the amplitude of the velocity anomaly when using the intensity weight \citep[e.g.][]{2005ApJ...631.1215W,2009A&A...506..377T,2010ApJ...709..458H}. Moreover, even two cross-correlation methods used in different instruments make $\lesssim$ 30 \% difference in the amplitude of the RM effect (T. Hirano in private communication). Hence more sophisticated modeling adapted to the details of measurement will be needed to obtain precise estimates practically. \citet{2010ApJ...709..458H} found that the radial velocity anomaly obtained by the cross-correlation method is larger than that predicted by the intensity weight method and that the bias tends to be larger as increasing the spin velocity. It makes detection of the spin effect easier. Detailed structures of planetary surface such as uneven molecular distribution \citep[e.g.][]{2010ApJ...719..341B} or winds may affect the PRV in principle. The PRV in transmission spectra \citep{2010Natur.465.1049S} might help to resolve a degeneracy between winds and spin \citep[see also][]{2007ApJ...669.1324S}. We postpone these problems until a next paper.

In conclusion, we have shown how the planet's rotation affects the PRV in the dayside spectra of exoplanets. We found that the spin effect of the solid rotation on the PRV is characterized by the projected angle between the orbital axis and the spin axis, $\lamp$. We also showed that the precise measurement of the PRV enable us to constrain the planetary obliquity via $\lamp$ and the spin period via $\Krot$. 

We are deeply grateful to Teruyuki Hirano for helpful discussion. We also thank an anonymous referee for a lot of constructive comments. HK is supported by a JSPS Grant-in-Aid for science fellows. This work is also supported by Grant-in-Aid for Scientific research from JSPS and from the MEXT (No. 22$\cdot$5467).


\begin{thebibliography}{50}

\bibitem[\protect\citeauthoryear{{Agnor}, {Canup}, \& {Levison}}{{Agnor}
  et~al.}{1999}]{1999Icar..142..219A}
{Agnor}, C.~B., {Canup}, R.~M.,  \& {Levison}, H.~F. 1999, Icarus, 142, 219

\bibitem[\protect\citeauthoryear{{Brogi} et~al.}{{Brogi} et~al.}{2012}]{brogi}
{Brogi}, M., {Snellen}, I.~A.~G., {de Kok}, R.~J., {Albrecht}, S., {Birkby},
  J.,  \& {de Mooij}, E.~J.~W. 2012, Nature, 486, 502

\bibitem[\protect\citeauthoryear{{Burrows} et~al.}{{Burrows}
  et~al.}{2010}]{2010ApJ...719..341B}
{Burrows}, A., {Rauscher}, E., {Spiegel}, D.~S.,  \& {Menou}, K. 2010, \apj,
  719, 341

\bibitem[\protect\citeauthoryear{{Chambers}}{{Chambers}}{2001}]{2001Icar..152..205C}
{Chambers}, J.~E. 2001, Icarus, 152, 205

\bibitem[\protect\citeauthoryear{{Cowan} \& {Agol}}{{Cowan} \&
  {Agol}}{2011}]{2011ApJ...726...82C}
{Cowan}, N.~B.,  \& {Agol}, E. 2011, \apj, 726, 82

\bibitem[\protect\citeauthoryear{{Cowan}, {Voigt}, \& {Abbot}}{{Cowan}
  et~al.}{2012}]{2012arXiv1205.5034C}
{Cowan}, N.~B., {Voigt}, A.,  \& {Abbot}, D.~S. 2012, arXiv:1205.5034

\bibitem[\protect\citeauthoryear{{Fossey}, {Waldmann}, \& {Kipping}}{{Fossey}
  et~al.}{2009}]{2009MNRAS.396L..16F}
{Fossey}, S.~J., {Waldmann}, I.~P.,  \& {Kipping}, D.~M. 2009, \mnras, 396, L16

\bibitem[\protect\citeauthoryear{{Fujii} \& {Kawahara}}{{Fujii} \&
  {Kawahara}}{2012}]{2012ApJ...755..101F}
{Fujii}, Y.,  \& {Kawahara}, H. 2012, \apj, 755, 101

\bibitem[\protect\citeauthoryear{{Gaudi} \& {Winn}}{{Gaudi} \&
  {Winn}}{2007}]{2007ApJ...655..550G}
{Gaudi}, B.~S.,  \& {Winn}, J.~N. 2007, \apj, 655, 550

\bibitem[\protect\citeauthoryear{{Heller}, {Leconte}, \& {Barnes}}{{Heller}
  et~al.}{2011}]{2011A&A...528A..27H}
{Heller}, R., {Leconte}, J.,  \& {Barnes}, R. 2011, \aap, 528, A27

\bibitem[\protect\citeauthoryear{{Hellier} et~al.}{{Hellier}
  et~al.}{2011}]{2011ApJ...730L..31H}
{Hellier}, C., {Anderson}, D.~R., {Collier-Cameron}, A., {Miller}, G.~R.~M.,
  {Queloz}, D., {Smalley}, B., {Southworth}, J.,  \& {Triaud}, A.~H.~M.~J.
  2011, \apjl, 730, L31

\bibitem[\protect\citeauthoryear{{Hirano} et~al.}{{Hirano}
  et~al.}{2010}]{2010ApJ...709..458H}
{Hirano}, T., {Suto}, Y., {Taruya}, A., {Narita}, N., {Sato}, B., {Johnson},
  J.~A.,  \& {Winn}, J.~N. 2010, \apj, 709, 458

\bibitem[\protect\citeauthoryear{{Kawahara} \& {Fujii}}{{Kawahara} \&
  {Fujii}}{2010}]{2010ApJ...720.1333K}
{Kawahara}, H.,  \& {Fujii}, Y. 2010, \apj, 720, 1333

\bibitem[\protect\citeauthoryear{{Kawahara} \& {Fujii}}{{Kawahara} \&
  {Fujii}}{2011}]{2011ApJ...739L..62K}
{Kawahara}, H.,  \& {Fujii}, Y. 2011, \apjl, 739, L62

\bibitem[\protect\citeauthoryear{{Kokubo} \& {Ida}}{{Kokubo} \&
  {Ida}}{2007}]{2007ApJ...671.2082K}
{Kokubo}, E.,  \& {Ida}, S. 2007, \apj, 671, 2082

\bibitem[\protect\citeauthoryear{{L{\'o}pez-Morales} \&
  {Seager}}{{L{\'o}pez-Morales} \& {Seager}}{2007}]{2007ApJ...667L.191L}
{L{\'o}pez-Morales}, M.,  \& {Seager}, S. 2007, \apjl, 667, L191

\bibitem[\protect\citeauthoryear{{Ohta}, {Taruya}, \& {Suto}}{{Ohta}
  et~al.}{2005}]{2005ApJ...622.1118O}
{Ohta}, Y., {Taruya}, A.,  \& {Suto}, Y. 2005, \apj, 622, 1118

\bibitem[\protect\citeauthoryear{{Pall{\'e}} et~al.}{{Pall{\'e}}
  et~al.}{2008}]{2008ApJ...676.1319P}
{Pall{\'e}}, E., {Ford}, E.~B., {Seager}, S.,
  {Monta{\~n}{\'e}s-Rodr{\'{\i}}guez}, P.,  \& {Vazquez}, M. 2008, \apj, 676,
  1319

\bibitem[\protect\citeauthoryear{{Queloz} et~al.}{{Queloz}
  et~al.}{2000}]{2000A&A...359L..13Q}
{Queloz}, D., {Eggenberger}, A., {Mayor}, M., {Perrier}, C., {Beuzit}, J.~L.,
  {Naef}, D., {Sivan}, J.~P.,  \& {Udry}, S. 2000, \aap, 359, L13

\bibitem[\protect\citeauthoryear{{Rodler}, {Lopez-Morales}, \&
  {Ribas}}{{Rodler} et~al.}{2012}]{2012ApJ...753L..25R}
{Rodler}, F., {Lopez-Morales}, M.,  \& {Ribas}, I. 2012, \apjl, 753, L25

\bibitem[\protect\citeauthoryear{{Smith} et~al.}{{Smith}
  et~al.}{2012}]{2012arXiv1203.6017S}
{Smith}, A.~M.~S., et~al. 2012, arXiv:1203.6017

\bibitem[\protect\citeauthoryear{{Snellen} et~al.}{{Snellen}
  et~al.}{2010}]{2010Natur.465.1049S}
{Snellen}, I.~A.~G., {de Kok}, R.~J., {de Mooij}, E.~J.~W.,  \& {Albrecht}, S.
  2010, \nat, 465, 1049

\bibitem[\protect\citeauthoryear{{Spiegel}, {Haiman}, \& {Gaudi}}{{Spiegel}
  et~al.}{2007}]{2007ApJ...669.1324S}
{Spiegel}, D.~S., {Haiman}, Z.,  \& {Gaudi}, B.~S. 2007, \apj, 669, 1324

\bibitem[\protect\citeauthoryear{{Triaud} et~al.}{{Triaud}
  et~al.}{2009}]{2009A&A...506..377T}
{Triaud}, A.~H.~M.~J., et~al. 2009, \aap, 506, 377

\bibitem[\protect\citeauthoryear{{Williams} \& {Kasting}}{{Williams} \&
  {Kasting}}{1997}]{1997Icar..129..254W}
{Williams}, D.~M.,  \& {Kasting}, J.~F. 1997, Icarus, 129, 254

\bibitem[\protect\citeauthoryear{{Williams} \& {Pollard}}{{Williams} \&
  {Pollard}}{2003}]{2003IJAsB...2....1W}
{Williams}, D.~M.,  \& {Pollard}, D. 2003, International Journal of
  Astrobiology, 2, 1

\bibitem[\protect\citeauthoryear{{Winn} et~al.}{{Winn}
  et~al.}{2005}]{2005ApJ...631.1215W}
{Winn}, J.~N., et~al. 2005, \apj, 631, 1215

\end{thebibliography}

\end{document}